\documentclass[aps,prb,twocolumn,superscriptaddress,showpacs]{revtex4}

\usepackage{graphicx}
\usepackage{bm}

\begin{document}

\title{Persistent spin current in mesoscopic ferrimagnetic spin
ring}
\author{Jing-Nuo Wu}
\affiliation{Department of Physics, National Taiwan Normal
University, Taipei, Taiwan}
\author{Ming-Che Chang}
\affiliation{Department of Physics, National Taiwan Normal University,
Taipei, Taiwan}
\author{Min-Fong Yang}
\affiliation{Department of Physics, Tunghai University, Taichung,
Taiwan}
\date{\today}
%%%%%%%%%%%%%%%%%%%%%%%%%%%%%%%%%%%%%%%%%%%%%%%%%%%%%%%%%%%%%%%%%%%

\begin{abstract}
Using a semiclassical approach, we study the persistent
magnetization current of a mesoscopic ferrimagnetic ring in a
nonuniform magnetic field. At zero temperature, there exists
persistent spin current because of the quantum fluctuation of
magnons, similar to the case of an antiferromagnetic spin ring. At
low temperature, the current shows activation behavior because of
the field-induced gap. At higher temperature, the magnitude of the
spin current is proportional to temperature $T$, similar to the
reported result of a ferromagnetic spin ring.
\end{abstract}
%%%%%%%%%%%%%%%%%%%%%%%%%%%%%%%%%%%%%%%%%%%%%%%%%%%%%%%%%%%%%%%%%%%
\maketitle

\section{Introduction}

Persistent charge current in a mesoscopic metal ring was
predicted\cite{Buttiker} and observed\cite{Chandrasekhar} a decade
ago. In such a ring threaded by a magnetic flux, if the phase
coherence length of electrons is larger than the size of the ring,
then the electrons can pick up an Aharonov-Bohm (AB) phase after
circling the ring once. Such a phase lag (or advance) would lead
to a persistent current, which is a periodic function of the
threaded magnetic flux,\cite{Byers} and can be detected via the
magnetic response of the (isolated) ring. The phase lag (or
advance) can also be of geometric origin (Berry
phase).\cite{Berry1984} It has been proposed that a Berry phase
can appear for an electron moving around the metal ring that
subject to a textured magnetic field (or
magnetization).\cite{Loss} This geometric phase, which depends
upon the solid angle associated with the textured magnetic field,
can lead to persistent charge and spin currents.\cite{Loss} A
similar geometric phase appears due to the spin-orbit interaction
in one-dimensional rings,\cite{Meir,MS} which is a manifestation
of the Aharonov-Casher (AC) effect.\cite{AC} More studies on the
persistent current related to the AC effect can be found in
Refs.~\onlinecite{BA,Choi,OR}.

With the advance of spintronics and quantum
computation,\cite{spintronics} it becomes more important to
understand the behavior of the spin current. Among these
investigations, spin transport in pure spin systems plays a
special role since there is no complication from charge degrees of
freedom. In a recent paper, using a semiclassical spin wave
analysis, Sch\"{u}tz {\it et al.} predicted the existence of
persistent spin current in a mesoscopic ferromagnetic (FM) spin
ring in a {\it nonuniform} magnetic field.\cite{Kollar01} The FM
spin ring being considered is a charge insulator with Heisenberg
spin interaction, and the spin current is carried by magnon
excitations. Similar to the case of charge transport in a metal
ring subject to a textured magnetic field,\cite{Loss} the magnon
in a mesoscopic FM spin ring acquires a geometric phase from the
(nonuniform) spin texture of the classical ground state. The
persistent current is found to be zero at temperature $T=0$, and
proportional to $T$ when $k_BT$ is larger than the field-induced
energy gap of the magnons.

Similar method has been applied to an antiferromagnetic (AFM) spin
ring with a Haldane gap.\cite{Kollar02} As compared with the FM
case, there are some subtleties in using the semiclassical method
in the AFM case. Due to the problem of infrared-diverging
magnetization, the spin-wave approach is not valid for AFM spin
chains with half-integer spins.\cite{Kollar02} That is the reason
why only the integer-spin cases are considered in
Ref.~\onlinecite{Kollar02}. Nonetheless, in the integer-spin case,
an additional staggered field in the direction of the classical
magnetization vectors still has to be introduced. Its value needs
to be determined self-consistently before quantitative predictions
can be made. The authors of Ref.~\onlinecite{Kollar02} find that,
unlike the case of the FM spin ring,  the persistent spin current
in an AFM spin ring can be nonzero at $T=0$ due to quantum
fluctuations. When the spin correlation length is much longer than
the size of the ring, the magnitude of the spin current exhibits
sawtooth variation with respect to the geometric phase, similar to
the case of persistent charge current in a metal ring. Recently,
the investigation has been extended to an anisotropic FM spin
ring,\cite{Bruno} a spin-1/2 AFM spin ring,\cite{Schmeltzer,Zhuo}
and an anisotropic AFM spin ring.\cite{Cheng}

In this paper, we study the persistent spin current in a
ferrimagnetic (FIM) spin ring with alternating spins $S^A$ and
$S^B$ under a textured magnetic field. Contrary to the AFM case,
the problem of infrared-diverging magnetization does not exist in
the present FIM case, no matter whether the constituent spins are
integer or half-integer.\cite{Brehmer,Pati,Wu,Yamamoto03} Thus the
self-consistently determined staggered field needs not be
introduced, and physical quantities can be calculated directly as
long as system parameters are known. We find that the FIM spin
ring can have either FM or AFM characteristics. For example, a
quantity proportional to $|S^A-S^B|$ plays a role similar to the
Haldane gap in the AFM spin ring. Moreover, a nonzero spin current
exists at $T=0$, again similar to the case of the AFM spin
ring.\cite{Kollar02} On the other hand, when the thermal energy is
higher than the field-induced gap, the magnitude of the spin
current is proportional to temperature $T$, similar to the case in
the FM spin ring.\cite{Kollar01}

This paper is organized as follows: We present the spin-wave
analysis in Sec.~II. The results of numerical calculations are
shown in Sec.~III, and Sec.~IV is the conclusion.

\section{theoretical analysis}

The Hamiltonian of the ferrimagnetic Heisenberg spin ring in a
nonuniform magnetic field ${\vec h}_{j}\equiv g\mu _{B}{\vec
B}(\vec{r}_j)$ is
\begin{equation}
H=J\sum_{j\in A\cup B} {\vec S}_{j}\cdot {\vec S}
_{j+1}-\sum_{j_1\in A,j_2\in B}\left( {\vec h}_{j_{1}} \cdot {\vec
S}_{j_{1}}^{A} +{\vec h}_{j_{2}}\cdot {\vec S}_{j_{2}}^{B}\right),
\end{equation}
where $J>0$, and the index $j$ refers to one of the alternating
$j_1,j_2$ sites. That is, sublattice-A can be labelled either by
$j$ (an odd integer) or $j_1=(j+1)/2$; sublattice-B can be
labelled by $j$ (an even integer) or $j_2=j/2$. There are even
number of lattice sites $N$. The length of the ring is $L$ and the
lattice spacing $a=L/N$. Periodic boundary condition ${\vec
S}_{j+N}={\vec S}_{j}$ is used. On the classical level, ${\vec
S}_{j}$ are replaced by classical vectors $S\hat{m}_{j}$. The
classical ground state $\{\hat{m}_{j}\}$ can be determined from
angular variations with respect to each ${\hat{m}_{j}}$, which
give
\begin{eqnarray}
& JS^{B}(\hat{m}_{j_2-1}+ \hat{m}_{j_2})
-{\vec h}_{j_1}+\lambda_{j_1}^A \hat{m}_{j_1}=0, \nonumber \\
& JS^{A}(\hat{m}_{j_1}+ \hat{m}_{j_1+1}) -{\vec
h}_{j_2}+\lambda_{j_2}^B \hat{m}_{j_2}=0,
\end{eqnarray}
where $\lambda_j$ are Lagrange multipliers. It shows that the
magnetization aligns parallel to the sum of external and exchange
field, as expected. If the Zeeman energy is much smaller than the
exchange energy between spins, then $\hat{m}_j^A$ and
$\hat{m}_{j+1}^B$ would be nearly antiparallel to each other, as
shown in Fig.~1(a). Moreover, due to nonzero magnetization in the
present FIM model, $\hat{m}_j^A$ would lie nearly along the
direction of $\vec{h}_j$, instead of nearly perpendicular to
$\vec{h}_j$ as in the AFM case\cite{Kollar02} (we take $S^A > S^B$
in this paper).

%%%%%%%%%%%%%%%%%%%%%%%%%%%%%%%%%%%%%%%%%%%%%%%%%%%%%%%%%%%%%%%%%%%%%
\begin{figure}
%\center
\includegraphics[width=3.2 in]{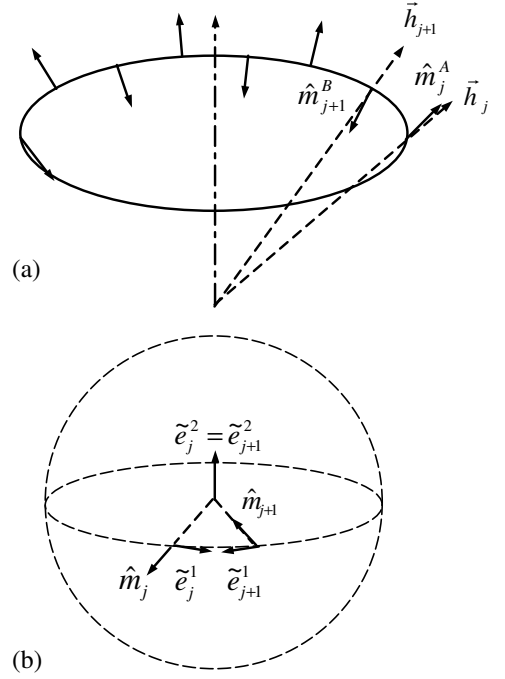}
\caption{(a) Classical spin configuration of a FIM spin ring in a
crown-shaped magnetic field. (b) Relative orientation for adjacent
local triads. The condition
$\tilde{e}_{j}^{2}=\tilde{e}_{j+1}^{2}=\hat{m}_{j}\times\hat{m}_{j+1}
/|\hat{m}_{j}\times\hat{m}_{j+1}|$ is imposed. The circle in the
middle is the equator of the sphere.}\label{fig1}
\end{figure}
%%%%%%%%%%%%%%%%%%%%%%%%%%%%%%%%%%%%%%%%%%%%%%%%%%%%%%%%%%%%%%%%%%%%%%

When quantum fluctuations are considered, following the treatment
introduced by  Sch\"{u}tz {\it et al.},\cite{Kollar01,Kollar02}
each spin operator is expanded using unit vectors that form an
orthogonal triad
$\left\{\hat{e}_{j}^{1},\hat{e}_{j}^{2},\hat{m}_{j}\right\}$,
\begin{equation}
{\vec S_{j}}=S_j^\parallel \hat{m}_{j}+\frac{1}{2}\left( S_{j}^+
\hat{e}_{j}^{-}+S_{j}^{-}\hat{e}_{j}^{+}\right),
\end{equation}
where $\hat{e}_{j}^{\pm }\equiv \hat{e}_{j}^{1}\pm
i\hat{e}_{j}^{2}$. We focus on systems with large spins (which
could be integer or half-integer), and introduce the
Holstein-Primakoff bosons $a_{j},b_{j}$,
\begin{eqnarray}
S_j^{A\parallel}&=&S^{A}-a_{j}^{\dagger}a_{j} \; ,  \quad
S_{j}^{A+}\cong \sqrt{ 2S^{A}}a_{j}^{\dagger} \; ; \nonumber
\\
S_j^{B\parallel}&=&S^{B}-b_{j}^{\dagger}b_{j} \; , \quad
S_{j}^{B+}\cong \sqrt{2S^{B}}b_{j}^{\dagger} \; .
\end{eqnarray}
Substituting these into Eq.~(1), we obtain
$H=H^{\parallel}+H^{\perp}+ H^{\prime}$ with
\begin{eqnarray}
H^{\parallel}&=&J\sum_{j\in A\cup B}S_j^{\parallel}
S_{j+1}^{\parallel}-\sum_{j\in A\cup B}S_j^\parallel
\hat{m}_j\cdot\vec{h}_j \; ,\\
H^{\perp}&=& J\sum_{j\in A\cup B}\vec{S}_j^{\perp}\cdot
\vec{S}_{j+1}^{\perp} \; , \\
H^{\prime}&=&\sum_{j_1\in A,j_2\in B} \left\{
\vec{S}_{j_1}^{A\perp}\cdot
\left[JS^B(\hat{m}_{j_2-1}+\hat{m}_{j_2})-{\vec
h}_{j_{1}}\right] \right. \nonumber\\
&& \left. +\vec{S}_{j_2}^{B\perp}\cdot
\left[JS^A(\hat{m}_{j_1}+\hat{m}_{j_1+1})-{\vec h}_{j_{2}}\right]
\right\} \nonumber\\
&&+O(\sqrt{S}).
\end{eqnarray}
$H^\parallel$ and $H^\perp$ are of the orders of $O(S^2)$ and
$O(S)$ respectively. To the order of $O(S)$, $H'$ is zero because
of Eq.~(2) and will be neglected in the following. In order to
simplify the Hamiltonian, we choose the local triads with the {\it
connection} shown in Fig.~1(b), in which
$\tilde{e}_{j}^{2}=\tilde{e}_{j+1}^{2}=\hat{m}_{j}\times
\hat{m}_{j+1}/|\hat{m}_{j}\times \hat{m}_{j+1}|$ (this is referred
to as a choice of gauge). Such {\it parallel transported} triads
are related to the original local triads by local rotations (gauge
transformations), $\hat{e}_{j}^{\pm }=e^{\pm i\omega_{j\rightarrow
j+1}}$ $\tilde{e}_{j}^{\pm }$, where $\omega_{j\rightarrow j+1}$
is the angle of rotation around $\hat{m}_j$ that takes
$\tilde{e}_j^2$ to $\hat{e}_j^2$.\cite{rotate}

Using the new triads in the Hamiltonian, we obtain
\begin{eqnarray}
H^\parallel = H_c &-& J\sum_{j}\hat{m}_{j}\cdot
\hat{m}_{j+1}\left[S^{A}\left(b_{j}^{\dagger}b_{j}
+b_{j+1}^{\dagger}b_{j+1}\right)\right.\nonumber
\\&+&\left.S^{B}\left(a_j^{\dagger}a_j
+a_{j+1}^{\dagger}a_{j+1}\right)\right] \nonumber \\
&+&\sum_{j_1,j_2} \left( h_{j_1}^A
a_{j_{1}}^{\dagger}a_{j_{1}}+h_{j_2}^B
b_{j_{2}}^{\dagger}b_{j_{2}} \right ),
\end{eqnarray}
where $H_c=JS^{A}S^{B}\sum_{j} \hat{m}_{j}\cdot
\hat{m}_{j+1}-\sum_{j_1,j_2} \left(
S^{A}h_{j_1}^{A}+S^{B}h_{j_2}^{B}\right)$ with $h_{j_1}^{A}
\equiv{\vec h}_{j_{1}}\cdot \hat{m}_{j_{1}}$ and $\
h_{j_2}^{B}\equiv{\vec h}_{j_{2}} \cdot \hat{m}_{j_{2}}$. Also,
\begin{eqnarray}
H^\perp &=&\frac{J}{2}\sqrt{S^{A}S^{B}} \nonumber \\
&\times&\sum_{j_1,j_2} \left\{ \left [ \left( 1+\hat{m}_{j_1}\cdot
\hat{m}_{j_2}\right) a_{j_1} b_{j_2}^{\dagger}e^{i(\omega
_{j_1\rightarrow j_2}-\omega
_{j_2\rightarrow j_1})} \right.\right. \nonumber \\
&+&\left( 1+\hat{m}_{j_2}\cdot \hat{m}_{j_1+1}\right)
a_{j_1+1}b_{j_2}^{\dagger}e^{i(\omega _{j_1+1\rightarrow
j_2}-\omega _{j_2\rightarrow j_1+1})} \nonumber \\
&-&\left(1-\hat{m}_{j_1}\cdot \hat{m}_{j_2}\right)
a_{j_1}^{\dagger}b_{j_2}^{\dagger}e^{-i(\omega _{j_1\rightarrow
j_2}+\omega _{j_2\rightarrow j_1})} \nonumber \\
&-&\left. \left( 1-\hat{m}_{j_2}\cdot \hat{m}_{j_1+1}\right)
a_{j_1+1}^{\dagger}b_{j_2}^{\dagger}e^{-i(\omega
_{j_1+1\rightarrow j_2}+\omega _{j_2\rightarrow j_1+1})} \right ]
\nonumber \\&+&\left. h.c. \right\}\;.
\end{eqnarray}
As long as the ring is not too small, neighboring spin vectors
would be nearly antiparallel to each other [as shown in Fig.~1(a)]
with $\hat{m}_{j}\cdot \hat{m} _{j+1}=-1+O(1/N)$. This further
simplifies the Hamiltonian to be
\begin{widetext}
\begin{eqnarray}
H &=& H_c+\sum_{j_1,j_2} \left[\left( 2JS^{B}+h_{j_1}^{A}\right)
a_{j_{1}}^{\dagger}a_{j_{1}} +\left( 2JS^{A}+\ h_{j_2}^{B}\right)
b_{j_{2}}^{\dagger}b_{j_{2}}\right]\nonumber \\
&-&J\sqrt{S^{A}S^{B}} \sum_{j_1,j_2}\left[
a_{j_1}^{\dagger}b_{j_2}^{\dagger}e^{-i(\omega _{j_1\rightarrow
j_2}+\omega _{j_2\rightarrow
j_1})}+a_{j_1+1}^{\dagger}b_{j_2}^{\dagger}e^{-i(\omega
_{j_1+1\rightarrow j_2}+\omega _{j_2\rightarrow
j_1+1})}+h.c.\right ].
\end{eqnarray}
\end{widetext}
The hopping of boson-$a$ from site-$j_1$ to site-$(j_1+1)$
acquires a phase $(\omega_{j_1\rightarrow j_2}+\omega
_{j_2\rightarrow j_1})-\left(\omega_{j_2\rightarrow j_1+1}+\omega
_{j_1+1\rightarrow j_2}\right)$. After circling around the ring
once, the boson gains a cumulative phase $\Omega
=\sum_{j=1}^N\left( -1\right)^{j+1}\left( \omega_{j\rightarrow
j+1}+\omega _{j+1\rightarrow j}\right)$, which is the holonomy
angle of the parallel transport and equals the solid angle
extended by the classical spin texture $\{\hat{m}_j
\}$.\cite{Berry} It can be shown that
\begin{equation}
\Omega={\rm Im}\log\prod_{j_1=j_2=1}^{N/2}\left
(\hat{e}_{j_1}^+\cdot \hat{e}_{j_2}^+\hat{e}_{j_2}^-\cdot
\hat{e}_{j_1+1}^- \right ),
\end{equation}
which is a gauge-invariant expression because the
$\hat{e}^+_j\hat{e}^-_j$ vectors always appear in pairs (note that
$\hat{e}^-_{N/2+1}=\hat{e}^-_1$). Therefore, this cumulative
geometric phase $\Omega$ is independent of the choices of the
local triads.

The phase factors in Eq.~(10) can be made implicit by merging with
the boson operators. Such new boson operators would satisfy the
twisted boundary conditions: $a_{j_1+N/2}=e^{i\Omega }a_{j_1}$,
$b_{j_2+N/2}=e^{-i\Omega }b_{j_2}$. With the help of the
transformations,
\begin{equation}
a_{k}=\sqrt{\frac{2}{N}}\sum_{j\in A}e^{-ikaj}a_{j}\; , \quad
b_{k}= \sqrt{\frac{2}{N}}\sum_{j\in B}e^{ikaj}b_{j}\;,
\end{equation}
in which
$$k_{n}=\frac{2\pi}{L}\left(n+\frac{\Omega}{2\pi}\right),
\quad n=0,1,2,...\frac{N}{2}-1 \; ,$$ to conform with the twisted
boundary condition, the Hamiltonian becomes
\begin{eqnarray}
H&=&H_c \nonumber\\
&&+\sum_{k}\left(2JS^B+h^A_q\right)a_{k+q}^\dagger
a_k+\left(2JS^A+h^B_q\right)b_{k+q}^\dagger b_k \nonumber\\
&& -2J\sqrt{S^A S^B}\sum_k\left(a_k b_k+a_k^\dagger
b_k^\dagger\right)\cos(ka),
\end{eqnarray}
where we have assumed that the applied magnetic field (as well as
$\{\hat{m}_j\}$) has only one Fourier component with momentum $q$
to simplify the expression. A crown-shaped magnetic field with
azimuthal symmetry has the $q=0$ component only. For convenience,
we consider the crown-shaped magnetic field below. For a large FIM
spin ring in a {\it weak} magnetic field, we also have $h^B_0
\cong -h_0^A \equiv -h_0$ [see Fig.~1(a)] with $h_0$ being
positive . With the help of the Bogoliubov transformation,
\begin{eqnarray}
a_{k}&=&\alpha_{k}\cosh \theta_{k}+\beta_{k}^{\dagger}
\sinh\theta_{k};\nonumber \\
b_{k}^{\dagger}&=&\alpha_{k}\sinh
\theta_{k}+\beta_{k}^{\dagger}\cosh \theta_{k},
\end{eqnarray}
and choosing
\begin{equation}
\tanh (2\theta _{k})=\frac{2\sqrt{S^{A}S^{B}}\cos (ka)}{
S^{A}+S^{B} },
\end{equation}
the Hamiltonian is finally diagonalized as
\begin{eqnarray}
H &=&\sum_{k}\left[ \epsilon_k^{-} \left(
\alpha_{k}^{\dagger}\alpha_{k}
+\frac{1}{2}\right)+\epsilon_k^{+}\left(
\beta_{k}^{\dagger}\beta_{k}+\frac{1}{2}
\right)\right] \nonumber \\
&&-\frac{NJS^{A}}{2}(1+\gamma),
\end{eqnarray}
where the two energy branches are
\begin{equation}
\epsilon_k^{\pm}=JS^{A}\left[ \sqrt{(1-\gamma)^2+4\gamma\sin^2
(ka)}\pm \left(1-\gamma\right) \right] \mp h_0,
\end{equation}
with $\gamma=S^{B}/S^{A}<1$. As mentioned before, contrast to the
AFM case,\cite{Kollar02} the staggered field needs not be
introduced in the present FIM case. Thus physical quantities can
be calculated directly as long as the system parameters are known.

Similar to the case of a FIM spin chain under a {\it uniform}
magnetic field,\cite{Maisinger} the magnons with energy
$\epsilon_k^{-}$ ($\epsilon_k^{+}$) in the present case correspond
to the ferromagnetic (antiferromagnetic) excitations. The energy
gaps of these two branches are $\epsilon_0^{-}=h_0$ and $
\epsilon_0^{+} =2JS^A(1-\gamma)-h_0$, respectively. That is, a gap
is induced by the applied field for the ferromagnetic excitations,
while the gap of the antiferromagnetic excitations is reduced by
the applied field. In the absence of external magnetic field, the
ferromagnetic branch $\epsilon_k^{-}$ becomes gapless with
quadratic $k$-dispersion at small $k$, which corresponds to the
Goldstone mode due to the spontaneously broken rotational
symmetry. Calculations using quantum Monte Carlo method yields
nearly the same curve for $\epsilon_k^{-}$, but $\epsilon_k^{+}$
is separated from $\epsilon_k^{-}$ with a larger ($k$-independent)
gap.\cite{Brehmer,Yamamoto03} Such a discrepancy is reduced when
the spins are larger and the semiclassical formalism works
better.\cite{Yamamoto00,Yamamoto03}

Since the magnons are non-interacting, it is straightforward to
obtain the free energy ($k_B\equiv 1$)
\begin{eqnarray}
F(\Omega ) &=&T\sum_{k} \ln \left[ 4\sinh \left( \frac{
\epsilon_k^{+}}{2T}\right) \sinh \left(
\frac{\epsilon_k^{-}}{2T}\right) \right]
\nonumber \\
&& -\frac{NJS^{A}}{2}\left( 1+\gamma \right). \quad\quad
\end{eqnarray}
Furthermore, it can be explicitly shown that the longitudinal
(gauge-invariant) spin current
\begin{equation}
I_{s}\equiv\left\langle {\hat m}_{j}\cdot {\vec I}_{j\rightarrow
j+1} \right\rangle=-\frac{\partial F(\Omega )}{\partial \Omega},
\end{equation}
similar to the relation for persistent charge current in a normal
metal ring, but with $\Omega$ replacing magnetic flux
$\phi$.\cite{Kollar01,Kollar02} From this relation we obtain the
magnetization current,
\begin{equation}
I_{m}=\frac{g\mu _{B}}{\hbar }I_{s} =-\frac{g\mu
_{B}}{L}\sum_{k,\;\alpha=\pm}
v_k^{\alpha}\left(n_k^{\alpha}+\frac{1}{2}\right),
\end{equation}
where
\begin{equation}
v_k^{\alpha}=\frac{1}{\hbar}\frac{\partial
\epsilon_k^{\alpha}}{\partial k}=\frac{2JS^{A} a}{\hbar} \cdot
\frac{ \gamma\sin(2ka)}{\sqrt{(1-\gamma)^2+4\gamma\sin^2 (ka)}}
\end{equation}
are the velocities of the magnons, and $n_k^{\alpha}=1/\left[ \exp
\left( \epsilon_k^{\alpha} /T\right) -1 \right]$ are the Bose
occupation numbers.

%%%%%%%%%%%%%%%%%%%%%%%%%%%%%%%%%%%%%%%%%%%%%%%%%%%%%%%%%%%%%%%%%%%%%
\begin{figure}
%\center
\includegraphics[width=3.2in]{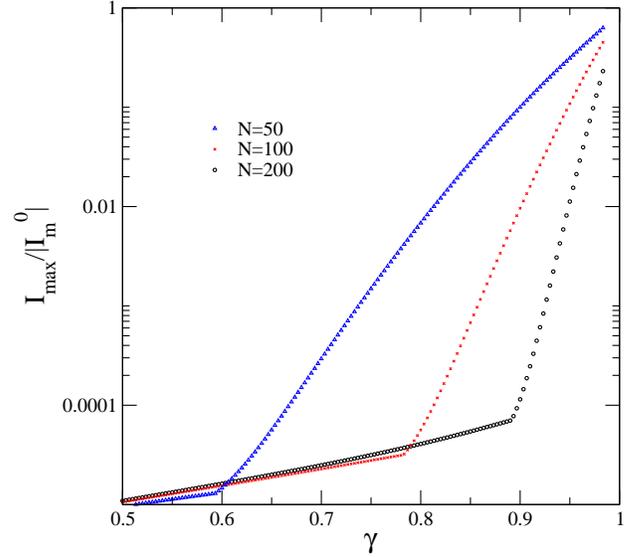}
\caption{Amplitude of the magnetization current at zero
temperature as a function of $\gamma$, plotted for three different
ring sizes.}\label{fig2}
\end{figure}
%%%%%%%%%%%%%%%%%%%%%%%%%%%%%%%%%%%%%%%%%%%%%%%%%%%%%%%%%%%%%%%%%%%%%%

\section{behavior of the persistent magnetization current}

For the FIM case, the magnetization current at $T=0$ is nonzero
even for vanishing magnon numbers (see Eq.~20), similar to the AFM
case,
\begin{equation}
I_{m}=I_{m}^{0} \sum_{k}
\frac{\gamma\sin(2ka)}{\sqrt{(1-\gamma)^2+4\gamma\sin^2 (ka)}} \;
,
\end{equation}
where $I_{m}^{0}\equiv-(2g\mu_{B}/\hbar)(JS^{A}/N)$. The magnon
velocity within the summation is a periodic function of $k$.
Therefore, after summing over the first Brillouin zone, the
current is zero if the $k$-points are distributed symmetrically
($\Omega=\pi$). For other values of $\Omega$, the summation is
nonzero but the magnitude of the magnetization current decreases
rapidly as $1-\gamma$ becomes larger. Comparing with the AFM
case,\cite{Kollar02} it can be seen that $\Delta \equiv \sqrt{
(1-\gamma)^2/4\gamma}$ plays a role similar to the Haldane gap,
and its inverse determines the scale of the spin correlation
length $\xi$. Therefore, when $\gamma$ is small enough such that
$\Delta\gg 2\pi/N$, the spin correlation length $\xi \ll L$; while
if $\gamma \simeq 1$, such that $\Delta\ll 2\pi/N$, we have $\xi
\gg L$. Therefore, by varying the ratio of the two different
spins, qualitatively different regimes, $\xi \gg L$ and $\xi \ll
L$, can be reached. In Fig.~2, the maximum amplitude of the
magnetization current $I_{\rm max}$ is plotted as a function of
$\gamma$. The functional form of $I_{\rm max}(\gamma)$ shows a
very clear crossover between these two different regimes.

%%%%%%%%%%%%%%%%%%%%%%%%%%%%%%%%%%%%%%%%%%%%%%%%%%%%%%%%%%%%%%%%%%%%%
\begin{figure}
%\center
\includegraphics[width=3.2in]{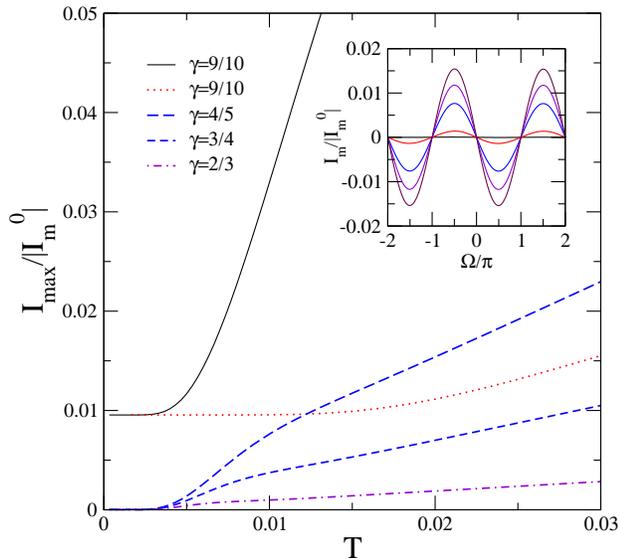}
\caption{Amplitude of the magnetization current as a function of
temperature $T$. Solid (dotted) line corresponds to a system
($N=100, \gamma=0.9$) in a magnetic field $h_0 / 2JS^A=$ 0.01
(0.05). Other curves with different $\gamma$ all have $h_0 /
2JS^A=0.01$. Inset: The variation of the magnetization current
with respect to the change of $\Omega$ and $T$ ($N=100,
\gamma=0.8$). The temperatures for the curves with the smallest
amplitude (indiscernible from the horizontal axis) to the largest
amplitude are $T/JS^A=$ 0, 0.005, 0.010, 0.015, 0.020, and
0.025.}\label{fig3}
\end{figure}
%%%%%%%%%%%%%%%%%%%%%%%%%%%%%%%%%%%%%%%%%%%%%%%%%%%%%%%%%%%%%%%%%%%%%%

Thermal energy could excite magnons and generate larger
magnetization current. Typical influence of the temperature on the
magnitude of the magnetization current can be seen in the inset of
Fig.~3. We have also studied the dependence of $I_{\rm max}$ on
the parameters $T$, $h_0$, and $\gamma$. From the magnon
dispersion relations in Eq.~(17), we expect activation behavior
for $I_{\rm max}(T)$ at low temperature $T<\epsilon_0^{-}=h_0$.
When the thermal energy is larger than the field-induced energy
gap $h_0$, $I_{\rm max}(T)$ should be proportional to $T$, similar
to the behavior of the persistent spin current in a FM spin
ring.\cite{Kollar01} These behaviors can be seen in Fig.~3. For
the upper two curves with $\gamma=0.9$, there is a significant
amount of spin current at zero temperature and the activation
behavior is implicit. Also, the persistent magnetization current
at very low $T$ is independent of the value of $h_0$ (see Eq.~22).

\section{conclusion}

We extend the work of Sch\"{u}tz {\it et
al.}\cite{Kollar01,Kollar02} and studied the persistent spin
current in a FIM spin ring. At $T=0$, the functional form of the
magnetization current $I_m(\gamma)$ shows distinctive behaviors
above/below a threshold value of $\gamma$, which depends on the
size of the ring. When thermal excitation can overcome the
field-induced energy gap $\epsilon_0^{-} = h_0$, the magnitude of
the spin current grows linearly with temperature $T$.

Both the persistent charge current in a metal ring and the
persistent spin current in a spin ring are related to geometric
phases (albeit of different origins). As in the case of the FM
spin ring, the induced electric voltage would be of the order of
$n$V, which poses a stringent experimental challenge. Recently,
several types of FIM spin-chain compounds have been
synthesized.\cite{Fujiwara,Escuer} To obtain a FIM spin chain with
$\gamma>0.8$, one might need to introduce rare earth
elements,\cite{large} or to fabricate a ring composed of magnetic
molecules with large spins.\cite{larges}

On the theoretical side, several questions remain open, such as
the generalizations to quasi-1D spin rings or spin rings with
itenerant electrons. The realistic effects of disorder,
interaction, or contact leads remain unknown in the spin ring
case. Further studies are desired.

\begin{acknowledgments}
J.N.W. and M.C.C. thank the support from the National Science
Council of Taiwan under Contract Nos. NSC 93-2112-M-003-009 and
NSC 94-2119-M-002-001. M.F.Y. acknowledges the support by the
National Science Council of Taiwan under NSC 93-2112-M-029-006.
\end{acknowledgments}


\begin{thebibliography}{99}

\bibitem{Buttiker}
M. Buttiker, Y. Imry, and R. Landauer, Phys. Lett. A \textbf{96},
365 (1983); H. F. Cheung, Y. Gefen, E. K. Riedel, and W. H. Shih,
Phys. Rev. B {\bf 37}, 6050 (1988); H. F. Cheung, Y. Gefen, and E.
K. Riedel, IBM J. Res. Dev. \textbf{32}, 359 (1988).

\bibitem{Chandrasekhar}
L. P. Levy, G. Dolan, J. Dunsmuir, and H. Bouchiat, Phys. Rev.
Lett. \textbf{64}, 2074 (1990); V. Chandrasekhar, R. A. Webb, M.
J. Brady, M. B. Ketchen, W. J. Gallagher, and A. Kleinsasser,
Phy. Rev. Lett. {\bf 67}, 3578 (1991).

\bibitem{Byers}
N. Byers and C. N. Yang, Phys. Rev. Lett. {\bf 7}, 46 (1961).

\bibitem{Berry1984}
M. V. Berry, Proc. Roy. Soc. London {\bf A 392,} 45 (1984).

\bibitem{Loss}
D. Loss, P. Goldbart and A. V. Balatsky, Phys. Rev. Lett. {\bf
65}, 1655 (1990); D. Loss and P. M. Goldbart, Phys. Rev. {\bf B
45}, 13~544 (1992).

\bibitem{Meir}
Y. Meir, Y. Gefen and O. Entin-Wohlman, Phys. Rev. Lett. {\bf 63},
798 (1989); O. Entin-Wohlman, Y. Gefen, Y. Meir and Y. Oreg, Phys.
Rev. {\bf B 45}, 11~890 (1992).

\bibitem{MS}
H. Mathur and A. D. Stone, Phys. Rev. Lett. {\bf 68}, 2964 (1992);
H. Mathur and A. D. Stone, Phys. Rev. {\bf B 44}, R10~
\begin{enumerate}
	\item 
\end{enumerate}
957 (1991).

\bibitem{AC}
Y. Aharonov and A. Casher, Phys. Rev. Lett. {\bf 53}, 319 (1984).

\bibitem{BA}
A. V. Balatsky and B. L. Altshuler, Phys. Rev. Lett. {\bf 70},
1678 (1993).

\bibitem{Choi}
M. Y. Choi, Phys. Rev. Lett. {\bf 71}, 2987 (1993).

\bibitem{OR}
S. Oh and C. M. Ryu, Phys. Rev. {\bf B 51}, 13~441 (1995).

\bibitem{spintronics}
D. Awschalom, N. Samarth, and D. Loss, {\it Semiconductor
Spintronics and Quantum Computation} (Springer, Berlin, 2002).

\bibitem{Kollar01}
F. Sch\"{u}tz, M. Kollar, and P. Kopietz, Phys. Rev. Lett.
\textbf{91}, 017205 (2003).

\bibitem{Kollar02}
F. Sch\"{u}tz, M. Kollar, and P. Kopietz, Phys. Rev. B
\textbf{69}, 035313 (2004).

\bibitem{Bruno}
P. Bruno, Phys. Rev. Lett. {\bf 93}, 247202 (2004); V.K. Dugaev,
P. Bruno, B. Canals, and C. Lacroix, Phys. Rev. B (in press);
cond-mat/0503013.

\bibitem{Schmeltzer}
D. Schmeltzer, A. Saxena, A. R. Bishop, and D. L. Smith,
cond-mat/0405659.

\bibitem{Zhuo}
W. Zhuo, X. Wang, and Y. Wang, cond-mat/0501693.

\bibitem{Cheng}
Y. Cheng, Y. Q. Li, and B. Chen, cond-mat/0505547.

\bibitem{Brehmer}
S. Brehmer, H.-J. Mikeska, S. Yamamoto, J. Phys.: Condens. Matter
\textbf{9}, 3921 (1997).

\bibitem{Pati}
S. K. Pati, S. Ramasesha, D. Sen, Phys. Rev. B {\bf 55}, 8894
(1997); J. Phys.: Condens. Matter \textbf{9}, 8707 (1997).

\bibitem{Wu}
C. Wu, B. Chen, Xi Dai, Yue Yu, and Z-B Su, Phys. Rev. B {\bf 60}, 1057 (1999).

\bibitem{Yamamoto03}
S. Yamamoto, cond-mat/0310004, also in {\it Recent Research
Developments in Physics}, Vol. 4 (Transworld Research Network,
Kerala, 2003), and references therein.
%It was shown that for a spin (1,1/2) FIM chain, the spin reduction
%using spin-wave analysis is approximately 0.305, while the correct
%value is about 0.207.

\bibitem{rotate}
If there are next-nearest-neighbor couplings in the Heisenberg
model, then the local rotation performed on a particular site
would depend on which neighbor it couples with.

\bibitem{Berry}
A. Shapere and F. Wilczek, {\it Geometric Phases in Physics}
(World Scientific, Singapore, 1989). In the discrete case, the
closed trajectory is connected by geodesics on the surface of the
unit sphere from $\hat{m}_j$ to $\hat{m}_{j+1}$.


\bibitem{Maisinger}
K. Maisinger, U. Schollw\"{o}ck, S. Brehmer, H.-J. Mikeska, and S.
Yamamoto, Phys. Rev. B \textbf{58}, R5908 (1998).

\bibitem{Yamamoto00}
S. Yamamoto, T. Fukui, and T. Sakai, Eur. Phys. J. B {\bf 15}, 211
(2000).

\bibitem{Fujiwara}
N. Fujiwara and M. Hagiwara, Solid State comm. {\bf 113}, 433
(2000).

\bibitem{Escuer}
A. S. Ovchinnikov, I. G. Bostrem, V. E. Sinitsyn, N. V. Baranov,
and K. Inoue, J. Phys.: Condens. Matter {\bf 13}, 5221 (2001); A.
S. Ovchinnikov, I. G. Bostrem, V. E. Sinitsyn, A. S. Boyarchenkov,
N. V. Baranov, and K. Inoue, J. Phys.: Condens. Matter {\bf 14},
8067 (2002).

\bibitem{large}
For example, Mn$^{2+}$ with $S=5/2$ and Cr$^{2+}$ (or Mn$^{3+}$)
with $S=2$ can give $\gamma=0.8$; Ho$^{3+}$ with $J=8$ and
Dy$^{3+}$ (or Er$^{3+}$) with $J=15/2$ can give $\gamma=0.937$.
For other possibilities, see Sec. 3 of Yamamoto's review paper in
Ref.~22.

\bibitem{larges}
D. Loss, D. P. DiVincenzo, and G. Grinstein, Phys. Rev. Lett. {\bf
69}, 3232 (1992); J. von Delft and C. L. Henley, Phys. Rev. Lett.
{\bf 69}, 3236 (1992); A. Garg, Europhys. Lett. {\bf 22}, 205
(1993).

\end{thebibliography}
\end{document}